\documentclass[onecolumn,usegraphicx,usenatbib]{mn2e}
\author[A. T. Potter \& C. A. Tout]
  {Adrian\,T.\,Potter and Christopher\,A.\,Tout\\
  University of Cambridge, Institute of Astronomy, The Observatories,
  Madingley Road, Cambridge CB3 0HA}
\title[White Dwarf Magnetic Fields in Strongly Interacting
  Binaries]{Magnetic Field Evolution of White Dwarfs in Strongly
  Interacting Binary Star Systems}
\date{\today}
\pagerange{\pageref{firstpage}--\pageref{lastpage}}
\pubyear{2009}
\begin{document}
\label{firstpage}
\maketitle

\begin{abstract}
The surface magnetic field strength of white dwarfs is observed to
vary from very little to around $10^9\,$G. Here we examine the
proposal that the strongest fields are generated by dynamo action
during the common envelope phase of strongly interacting stars that
leads to binary systems containing at least one white dwarf. The
resulting magnetic field depends strongly on the electrical
conductivity of the white dwarf, the lifetime of the convective
envelope and the variability of the magnetic dynamo. We assess the
various energy sources available and estimate necessary lifetimes of
the common envelope. In the case of a dynamo that leads a randomly
oriented magnetic field we find that the induced field is confined to
a thin boundary layer at the surface of the white dwarf. This then
decays away rapidly upon dispersal of the common envelope. The
residual field is typically less than $10^{-8}$ times the strength of
the external field. Only in the case where there is some preferential
direction to the dynamo-generated field can an induced field, that
avoids rapid decay, be produced.
We show that a surface field of magnitude a few\,per cent of the
external field may be produced after a few Myr. In this case the
residual field strength is roughly proportional to the lifetime of the
dynamo activity.

\end{abstract}

\section{Introduction}
\label{intro}

Surveys of the galactic white dwarf (WD) population have discovered
magnetic field strengths ranging up to about $10^9\,\rm{G}$
\citep{schmidt2003}. Typically WDs fall into two categories, those
with field strengths of the order $10^6\,\rm{G}$ or higher and those
with fields weaker than around $10^5\,\rm{G}$. We focus here on highly
magnetic WDs (hereinafter MWDs) with field strengths greater than
$10^6\,\rm{G}$. \citet{landstreet1971} proposed a fossil field
mechanism based on the evolution of Ap/Bp stars but it
is difficult to 
argue that such a field can survive stellar evolution in which all
parts of the star have passed through convective phases.
Here we build on the
proposal of \citet{tout2008} that the origin of such fields lies in
the interaction between the WD and its companion star in a binary
system.

This assertion is based on observations from the SDSS that
approximately $10$\,per cent of isolated WDs are highly magnetic
\citep{liebert2005} as are $25$\,per cent in cataclysmic
variables \citep{wickramasinghe2000} while there are none to be found
in wide detached binary systems.  Of the 1\,253 binary systems
comprising a WD and non-degenerate M-dwarf star surveyed in the SDSS
Data Release Five \citep{silvestri2007} none have been identified
with magnetic fields greater than the detection limit of about
$3\,\rm{MG}$. The relatively high occurrence of MWDs in strongly
interacting binaries compared with elsewhere suggests that the
generation of the strong magnetic fields is likely the result of the
interaction and subsequent evolution of the binary system. The MWDs
observed in isolated systems may be explained by either the total
disruption of the companion star during unstable mass transfer or the
coalescence of the MWD and the core of its companion following loss of
sufficient orbital energy to the common envelope (hereinafter CE) or
via gravitational radiation.

When a giant with a degenerate core expands beyond its Roche Lobe
mass transfer may proceed on a dynamical time scale.  A dense,
typically main-sequence star,
companion cannot accrete the overflowing material fast enough so that
it instead swells up to form a giant
CE. As a result of energy transfer and angular momentum to the CE during orbital decay of
the dense companion and the remnant core, strong differential
rotation is established within the envelope. Also, owing to its size
and thermal characteristics together with the nuclear energy source at
the core, the CE is expected to be largely
convective. In the mechanism proposed by Tout et al. this is expected
to drive strong dynamo action giving rise to powerful magnetic fields
\citep{regos1995,tout1992}. If sufficiently strong
dynamo action occurs in the CE then comparable magnetic fields may be
induced in the degenerate core (DC) that will evolve into a WD once
the envelope has been removed.  We show here that strong surface fields can result
from CE evolution. The strength of such fields is highly dependent on
the electrical conductivity of DC, the lifetime of the CE and the
variability of the magnetic dynamo.

In section 2 we outline the various sources of energy in the binary/CE system and their relation to the energy requirements of the magnetic dynamo. In section 3 we discuss the governing equations of the system and derive the form of the magnetic field for a general spatially and temporally varying magnetic diffusivity $\eta({\textbf{\em{r}}},t)$ via the static induction equation. Then in section 4 we give an overview of the numerical methods we have used to solve the various stages of the problem. In section 5 we present our results and we discuss these and conclude in section 6.

\section{CE Evolution and energy constraints}
\label{energy}

In the absence of detailed hydrodynamic properties, the most favoured
models for common envelope evolution are the so-called $\alpha$
\citep{webbink1976,livio1988} and $\gamma$
\citep{paczynski1967,nelemans2000} prescriptions which use a one
dimensional parametrization of the transfer of energy and angular
momentum respectively between the binary orbit and the envelope. We
consider the energy content of a typical envelope compared to the
energy necessary to generate the desired magnetic fields.

The primary sources of energy in the CE are the orbital energy of the
binary system itself and the gravitational binding energy of the
envelope. In the $\alpha$ prescription, energy is transferred from the
orbit to the envelope and this leads to the expulsion of the envelope
and decay of the orbital separation. This is regarded as the most
likely origin of cataclysmic variables
\citep{paczynski1976,meyer1979}. Other potential sources are the
thermal energy content of the envelope and its rotational
energy. However for a virialised cloud, these energies are small
compared to the other energy sources. In some systems recombination
energy can be similar in magnitude to the binding energy so may become
important \citep{webbink2008}.

Let $W,$ $O$ and $M$ be the binding, orbital and magnetic energy content of the binary/CE system respectively.

\begin{eqnarray}
W&=&\eta_{w}\frac{GM_{\rm{env}}M_{\rm{T}}}{R_{\rm{env}}},\\
O&=&\frac{GM_{\rm{c}}M_{\rm{2}}}{a},\\
{\rm and}\\
M&=&\frac{B^2}{2\mu_{\rm{0}}}(2\pi a)(\pi r_{\rm{I}}^2),
\end{eqnarray}

\noindent where $M_{\rm{env}}$ is the mass of the CE with radius
$R_{\rm{env}}$, $M_{\rm{c}}$ is the mass of the degenerate core,
$M_{\rm{2}}$ is the mass of the dense companion, $\eta_{\rm{w}}$ is a
constant of order unity that might be estimated from stellar
models. The total mass of the system is
$M_{\rm{T}}=M_{\rm{env}}+M_{\rm{c}}+M_{\rm{2}}$ and $a$ is the final
orbital separation. As we would expect, the total available
energy is less than the energy required to produce a field of the
desired strength throughout the cloud. However, given that we
anticipate that the field is generated by dynamo action, the strongest
fields occur where the hydrodynamic motions of the CE are most
strongly perturbed. We imagine this region to be a torus of the
orbital radius $a$ and cross-sectional radius $r_{\rm{I}}$ which is a
few times $\max(R_{\rm{2}},R_{\rm{c}})$ where $R_{\rm{2}}$ is the
radius of the dense companion and $R_{\rm{c}}$ is the radius of the core of the
giant. The energy of the cloud is then approximated by

\begin{eqnarray}
W&=&4.6\times10^{38}\,\eta_{\rm{w}}\left(\frac{M_{\rm{env}}}{M_{\odot}}\right)\left(\frac{M_{\rm{T}}}{2.6\,M_{\odot}}\right)\left(\frac{R_{\rm{env}}}{10\,\rm{au}}\right)^{-1}\rm{J},\\
O&=&1.0\times10^{41}\,\left(\frac{M_{\rm{c}}}{0.6\,M_{\odot}}\right)\left(\frac{M_{\rm{2}}}{M_{\odot}}\right)\left(\frac{a}{0.01\,\rm{au}}\right)^{-1}\rm{J},\\
{\rm and}\\
M&=&5.7\times10^{39}\,\left(\frac{B}{10^7\,\rm{G}}\right)^2\left(\frac{a}{0.01\,\rm{au}}\right)\left(\frac{r_{\rm{I}}}{R_{\odot}}\right)^2\rm{J}\label{me}.
\end{eqnarray}

Most of the energy may be derived from the orbital decay of the binary
within the CE. The value of $a$ we have taken is for a typical
separation with orbital period of $1\,$d and thus represents the final
separation of the system. As the orbit decays, the increase in orbital
velocity results in stronger local perturbations to the CE and in an
increase in dynamo activity that produces stronger magnetic fields
provided the CE is still sufficiently dense around the stellar
cores. Therefore it is more appropriate to consider the late stage of
CE evolution from the point of view of magnetic dynamos.

For the AM Herculis system $M_{\rm{c}}=0.78\,M_{\odot}$
\citep{gansicke2006}, $M_{\rm{2}}=0.37\,M_{\odot}$
\citep{southwell1995}, $a=1.1\,R_\odot$ \citep{kafka2005} and
$B=1.4\times10^7\,\rm{G}$ \citep{wickramasinghe1985}. Taking $M=O$ gives
$r_{\rm{I}}=2.1\left({B}/{10^9\,\rm{G}}\right)^{-1}\,R_{\odot}$. This
represents the limiting radius in which a field of sufficient strength
were generated given the total energy available. This is far larger
than the radius of the DC and somewhat larger than the second star and
so energetically there is nothing to prevent the necessary field from
being generated.

\section{Governing equations for magnetic field evolution}
\label{equations}

Consider the system evolving according to the static induction
equation for a general isotropic magnetic diffusivity field,
$\eta({\textbf{\em{r}}},t)$, related to the electrical conductivity,
$\sigma$ by $\eta=\frac{c^2}{4\pi\sigma}$.  The DC is embedded in an
infinite, uniform, vertical, time dependent magnetic field. This is a
sensible approximation provided the length scale for variation of the
external field defined by the dynamo action in the CE is sufficiently
large compared to the radius of the DC. We expect the DC to be
  spherically symmetric and we further assume the external field to be
  locally axisymmetric at the surface of the DC. In reality the exact
form of the imposed field is uncertain and is likely to support a
complex geometry. This is supported by spectropolarimetric
observations of WD magnetic field morphologies \citep{valyavin2006}. We
consider what effect this might have later on.

Except where explicitly stated otherwise, we are working in spherical
polar coordinates $(r,\theta, \phi)$. Outside of the DC
($r>r_{\rm{c}}$) we require

\begin{equation}
{\mathbf{\nabla\times}}{\textbf{\em{B}}}={\textbf{\em{0}}}
\label{external}
\end{equation}
\noindent and
\begin{equation}
{\textbf{\em{B}}}\to B_z(t){\textbf{\em{e}}}_z \quad\rm{as}\quad r\to\infty.
\end{equation}

\noindent Equation~(\ref{external}) is satisfied locally
  around the DC because the magnetic field is a superposition of the curl-free
  imposed field and the dipole field produced by the DC. In the global
  field we expect this condition to be broken to the motions of the
  CE. Inside the DC the magnetic field evolves according to the MHD
induction equation with stationary fluid,

\begin{equation}
\label{induction}
\frac{d{\textbf{\em{B}}}}{dt}={\mathbf{\nabla\times}}(\eta{\mathbf{\nabla\times}}{\textbf{\em{B}}}).
\end{equation}

\noindent We have ignored the term
${\mathbf{\nabla\times}}({\textbf{\em{u}}}\times{\textbf{\em{B}}})$, assuming
that this term, that results from fluid motions within the DC, is
dominated by the diffusive term. If this term were large then we would
expect WDs to support dynamo action.  This is not supported by the
statistics presented in section \ref{intro}. The magnetic field must
be continuous at $r=r_{\rm{c}}$. We proceed by decomposing the
magnetic field into poloidal and toroidal parts

\begin{equation}
{\textbf{\em{B}}}={\mathbf{\nabla\times}}(T{\textbf{\em{r}}})+{\mathbf{\nabla\times}}({\mathbf{\nabla\times}}(S{\textbf{\em{r}}})),
\end{equation}

\noindent where the first term on the right is the toroidal part of
the magnetic field and the second term is the poloidal part. In the
case of the magnetic field external to the DC, equation
(\ref{external}) simplifies to

\begin{equation}
{\mathcal {L}} ^2T=0, \quad\rm{and}\quad {\mathcal{L}}^2(\nabla^2S)=0,
\end{equation}
\noindent where
\begin{equation}
{\mathcal{L}}^2=-\left\{\frac{1}{\sin\theta}\frac{\partial}{\partial\theta}\left[\sin\theta\frac{\partial}{\partial\theta}\right]+\frac{1}{\sin^2\theta}\frac{\partial^2}{\partial\phi^2}\right\}.
\end{equation}

\noindent Taking $S=\sum_{l=1}^{\infty}S_l(r,t)P_l(\cos\theta)$
(similarly for $T$) and given
${\mathcal{L}}^2P_l(\cos\theta)=l(l+1)P_l(\cos\theta)$ this implies

\begin{equation}
T=0,\quad  \nabla^2S=0.
\end{equation}

\noindent The condition ${\textbf{\em{B}}}\rightarrow B_z{\textbf{\em{e}}}_z$ is
equivalent to $S\rightarrow\frac{1}{2}B_zr \cos\theta$. The solutions
of $\nabla^2S=0$ give $S$ in terms of spherical harmonics. In the
far-field the only mode is that corresponding to $l=1$. The other
modes decay exponentially without some mechanism to regenerate
them. Thus we take $S$ outside of the DC to be of the form

\begin{equation}
S=\left(S_0(t)r^{-2} + \frac{1}{2}B_z(t)r \right)\cos\theta.
\end{equation}

\noindent where $B_z(t)$ is the external field arising from the dynamo
activity in the CE. Now consider the field inside the DC. Provided the
magnetic diffusivity is spherically symmetric we can again decompose
${\textbf{\em{B}}}$ into poloidal and toroidal parts to see that $S$ and
$T$ evolve according to

\begin{equation}
\dot{S}=\eta(r,t)\nabla^2S \quad{\rm{and}}\quad \dot{T}=\eta (r,t)\nabla^2T.
\end{equation}

\noindent If we assume that we may write the diffusivity in the
self-similar form $\eta(r,t)=\eta_r(r)\eta_t(t)$ then the equation is
completely separable. So suppose we write
$T({\textbf{\em{r}}},t)=U(t)V({\textbf{\em{r}}})$ then we find that the two
functions satisfy

\begin{equation}
\frac{\dot{U}}{\eta_t U}=\frac{\eta_r\nabla^2 V}{V}=-\lambda^2, \quad \rm{where}\,\lambda\,\rm{is\, a\, complex\, constant}
\end{equation}
\noindent so that
\begin{equation}
\label{17}
U=\exp\left(-\lambda^2\int\eta_t(t)dt\right)
\end{equation}
\noindent and
\begin{equation}
\nabla^2V+\frac{\lambda^2}{\eta_r}V=0\label{Helm}.
\end{equation}

In the case of spatially constant diffusivity, equation~(\ref{Helm})
is simply Helmholtz' equation. Choosing solutions which are bounded as
$r\rightarrow0$ gives solutions of the form $V\propto j_1(a
r)\cos\theta$ where $j_i$ represents the $i^{th}$ spherical Bessel
function of the first kind and $a=\sqrt{\frac{\lambda^2}{\eta_r}}$. In
order to satisfy continuity in the magnetic field at $r=r_{\rm{c}}$ we
must have $j_1(ar_{\rm{c}})=0$. This gives a real value of $a$ and so
$\lambda$ must be real. So with no way to replenish the toroidal
field, $T\rightarrow0$ exponentially by equation~(\ref{17}).

It should be noted that although we have assumed that all higher order
spherical harmonics, azimuthal modes of the toroidal magnetic field
decay exponentially we have not presented here additional
consideration to the relative decay times. Given the vertical form of
the imposed external field, the dipole mode is the only one induced by
the CE and therefore the only mode we focus on. A similar analysis may
be performed for other values of $l$ but these modes are only
important if they are produced by the external field.

The equation for $S$ can be treated in a very similar way to $T$. For $r<r_{\rm{c}}$ we write 

\begin{equation}
\label{S}
S=\int_{-\infty}^\infty R(r;\gamma)\exp\left(\rm{i} \gamma H\right) \cos\theta \,d\gamma.
\end{equation}

\noindent By taking $H=\int\eta_t(t)dt$ the system is reduced to a
Fourier analysis. It is possible to use different transform methods
with different choices of the parameter $\gamma$. However, because we
shall consider an oscillating external field, this is the natural
choice. In this form the function is still separable and $R$ solves
the equation

\begin{equation}
\nabla^2(R \cos\theta)=\frac{{\rm{i}} \gamma}{\eta_r} R \cos \theta
\end{equation}
\noindent so that
\begin{equation}
\frac{d^2R}{dr^2}+\frac{2}{r}\frac{dR}{dr}-(\frac{{\rm{i}}\gamma}{\eta_r}+\frac{2}{r^2})R=0.
\label{Helmholtz}
\end{equation}

\noindent In order to enforce continuity at the boundary we must have
$S\,\rm{and}\,\frac{\partial S}{\partial r}$ continuous at
$r=r_{\rm{c}}$ for all $\theta$. These are equivalent to

\begin{equation}
\label{rearrange1}
\int_{-\infty}^{\infty}\frac{\partial R(r;\gamma)}{\partial r}\bigg|_{r=r_{\rm{c}}}\exp\left({\rm{i}}\gamma H\right)\,d\gamma=-\frac{2S_0(t)}{r_{\rm{c}}^3}+\frac{1}{2}B_z(t)\\
\end{equation}
\noindent and
\begin{equation}
\label{rearrange2}
\int_{-\infty}^{\infty}R(r_{\rm{c}},\gamma)\exp({\rm{i}}\gamma H)\,d\gamma=\frac{S_0(t)}{r_{\rm{c}}^2}+\frac{1}{2}B_z(t)r_{\rm{c}}.
\end{equation}

\noindent By taking $\frac{1}{3}((\ref{rearrange2})-r_{\rm{c}}.(\ref{rearrange1}))$ and $\frac{4}{3}((\ref{rearrange2})/r_{\rm{c}}+\frac{1}{2}(\ref{rearrange1}))$ these become

\begin{equation}
\frac{1}{3}\int_{-\infty}^{\infty}\left(R(r_{\rm{c}},\gamma)-\frac{\partial R(r;\gamma)}{\partial r}\bigg|_{r=r_{\rm{c}}}.r_{\rm{c}}\right) \exp({\rm{i}}\gamma H)\,d\gamma=\frac{S_0(H)}{r_{\rm{c}}^2}\label{dipole}\\
\end{equation}
\noindent and
\begin{equation}
\frac{4}{3}\int_{-\infty}^{\infty}\left(\frac{R(r_{\rm{c}},\gamma)}{r_{\rm{c}}}+\frac{1}{2}\frac{\partial R(r;\gamma)}{\partial r}\bigg|_{r=r_{\rm{c}}}\right) \exp({\rm{i}}\gamma H)\,d\gamma=B_z(H)\label{boundary}.
\end{equation}

\noindent We take ${\textbf{\em{B}}}(H(t))$ as given and so focus for the
moment on the second of these equations. This allows us to solve for
$S$ and then we can derive the magnetic field within the DC. So if the
transform of $B_z(H)$ is $\hat{B}_z(\gamma)$ then we may rewrite the
previous equation as

\begin{equation}
\label{boundary2}
\hat{B}_z(\gamma)=\frac{4}{3}\left(\frac{R(r_{\rm{c}};\gamma)}{r_{\rm{c}}}+\frac{1}{2}\frac{\partial R(r;\gamma)}{\partial r}\bigg|_{r=r_{\rm{c}}}\right).
\end{equation}

\noindent So given the condition $R\to 0$ as $r\to 0$ and this boundary condition we can now fully determine $R$. This in turn fully solves the internal (and external) dipole field of the DC. In the case $T=0$, $S=Q(t)R(r)\cos\theta$, the internal magnetic field (for a particular $\gamma$) is given by

\begin{equation}
\label{field}
{\textbf{\em{B}}}=Q(t)\left(2\frac{R}{r}\cos\theta\,{\textbf{\em{e}}}_r-\left(\frac{R}{r}+\frac{\partial R}{\partial r}\right)\sin\theta\,{\textbf{\em{e}}}_{\theta}\right).
\end{equation}

\noindent In theory we may solve this system exactly for any imposed
field. However, owing to the complexity of the external field and its
Fourier transform under the change of variable $t\to H(t)$, numerical
computations become extremely difficult. Thus, for the most part, we
focus on single modes given by some specific $\gamma$. This is
ultimately justified by the slow variation of $\eta$ during the
lifetime of the CE.

In the case where the induced magnetic field is not entirely dipolar,
the final term in equation~(\ref{Helmholtz}) is modified by some
factor for each $l$. In the case of uniform diffusivity this gives
spherical Bessel functions of varying order. The forms of these
functions in this parameter regime are actually very similar to the
first order function considered in section \ref{uniform} so we do not
expect including the higher modes to affect the qualitative results
significantly for the overall strength and radial variation of the
internal field. In addition, given the uncertainties regarding the
geometry of the magnetic field of the CE, inclusion of higher order
harmonics is unlikely to give any better insight.

\section{Numerical Methods}

We calculated the spatial form of the magnetic field in the presence
of an oscillating external field with a sixth order adaptive step
Runge-Kutta algorithm \citep{press1992} according to equation
(\ref{field}). The equation was integrated from the centre out to the
surface with boundary conditions $R(0)=0$ and a small arbitrary value
of $R'(0)$. The solution grows by several thousand orders of magnitude
between $r=0$ and $r=r_{\rm{c}}$. To cope with this variation, we
included a subroutine to rescale the solution whenever $R(r)$ or its
derivatives exceeded some maximum value, typically $10^{30}$. Once a
solution had been found the boundary condition at $r=r_{\rm{c}}$,
given by equation~(\ref{boundary}), was matched by scaling the entire
function. Such arbitrary rescalings are valid because of the homogeneous
nature of the governing ODE and the zero boundary condition at
$r=0$. In order to calculate the diffusivity field we employed the
approximation of \citet{wendell1987} for the
electrical conductivity. In addition we needed to calculate the
opacity for the given temperature and density. This was done with the
data tables and subroutines of the STARS stellar evolution code
\citep{eggleton1973,pols1995}.  In all simulations we have
used a CO~WD composition for the DC.

The time evolution of the magnetic field following the dispersal of the external field was calculated with a first order Euler finite step method. The code was run multiple times with a variety of step sizes and spatial resolutions with no significant variation in the solution between runs. The induction equation was simplified by the relations

\begin{equation}
\tilde{R}(r)=R(r)+\frac{r}{2}\frac{\partial R}{\partial r},
\end{equation}
\noindent and
\begin{equation}
R(r)=\frac{1}{r^2}\int^r_02r\tilde{R}(r)dr. \label{Q_1}
\end{equation}

\noindent This removes the spherical geometry of the induction equation, reducing it to the form

\begin{equation}
\frac{\partial \tilde{R}}{\partial t}=\eta\frac{\partial^2\tilde{R}}{\partial r^2},
\end{equation}

\noindent which has a zero boundary condition at $r=r_{\rm{c}}$ in the
absence of an external field and is ultimately easier to work
with. The function $R$ and subsequently ${\textbf{\em{B}}}$ are then
recovered from $\tilde{R}$ with the relation in equation~(\ref{Q_1})
numerically evaluated with the trapezoidal rule.  The routine does not
take into account DC cooling. This is unlikely to affect the evolution
of the field during the CE phase and shortly after when the field is
evolving rapidly. In the late time evolution however when the field
has finished its radial redistribution and is decaying purely
exponentially, \citet{wendell1987} showed that the
cooling significantly increases the decay timescale and the magnetic
field becomes essentially frozen into the dwarf. Therefore we may take
the final value for the field strength shortly after it has reached a
state of exponential decay. This is typically around $10^7\,$yr after
the dispersion of the CE.

\section{Results}
\label{uniform}

The magnetic field produced by the CE is heavily dependent on several
key factors. Although the magnetic field of the DC and the CE is
continuous at the surface, the total magnetic flux able to penetrate
the DC depends on the lifetime of the CE. Once the CE has
dispersed, the field continues to migrate inwards and the surface
field strength to decay. The degree to which this happens depends on
how much of the field is able to penetrate the DC while the surface
field is still maintained. This requires a suitable treatment of the
radial conductivity profile of the DC. As the density increases
towards the centre of the DC the conductivity rises rapidly,
inhibiting further diffusion of the field. This results in the
confinement of the field to around the outer 10\,per cent of the DC
by radius.

The structure and orientation of the magnetic field of the CE is also of critical importance. In a convectively driven magnetic dynamo we might expect the orientation of the field to change rapidly. The frequency of the changes has important consequences for the field. Given the geometry of the system driving the fluid motions which in turn give rise to the magnetic dynamo we might also anticipate a preferred direction to the orientation of the field. A field that is maintained in a single direction produces a WD field several orders in magnitude stronger than one in which the orientation varies rapidly and with random orientation.

\subsection{Consequences of CE lifetime}

The lifetime of the CE has a complicated relationship to the original
orbital separation of the binary system, the size of the envelope and
the properties of the degenerate core and its companion. This is to be
expected because of the wide range of magnetic field strengths in
MWDs. Here we consider how quickly a field may be built up in the DC
as a result of an applied uniform constant vertical magnetic field
outside. This mechanism produces stronger magnetic fields than one
with varying orientation but it is the best case scenario and places
an upper bound on how much field might be retained.  While the MHD
properties of CEs are not understood, it is difficult to say to what
degree there may be a preferred orientation for the magnetic field so
the true physical system may resemble anything in between a
uni-directional field and one that reorients itself randomly in any
direction.

\begin{figure}
\centering
\includegraphics[width=0.8\textwidth]{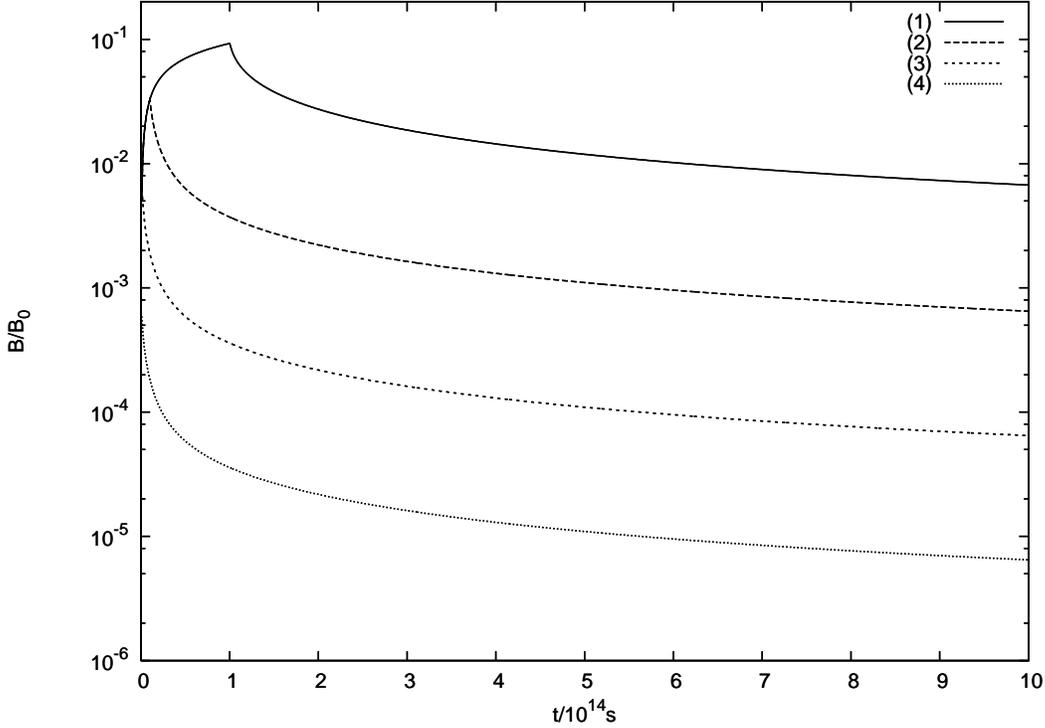}
\caption{The radial magnetic field strength at $r=0.99\,r_{\rm{c}}$
  that results from a constant, uniform, vertical magnetic field
  applied to a DC for (1) $10^{11}\,\rm{s}$, (2) $10^{12}\,\rm{s}$,
  (3) $10^{13}\,\rm{s}$, (4) $10^{14}\,\rm{s}$. Note the peak of the
  field strength at the end of the phase where the external field is
  applied. After the removal of the external field there is a period
  of rapid decay lasting around $2\times 10^{14}\,\rm{s}$ after which
  the field decays much more slowly on a timescale of around
  $10^{15}\,\rm{s}$.  The strength of the field is
  roughly proportional to the lifetime of the external field.}
\label{applied}
\end{figure}

\begin{figure}
\centering
\includegraphics[width=0.8\textwidth]{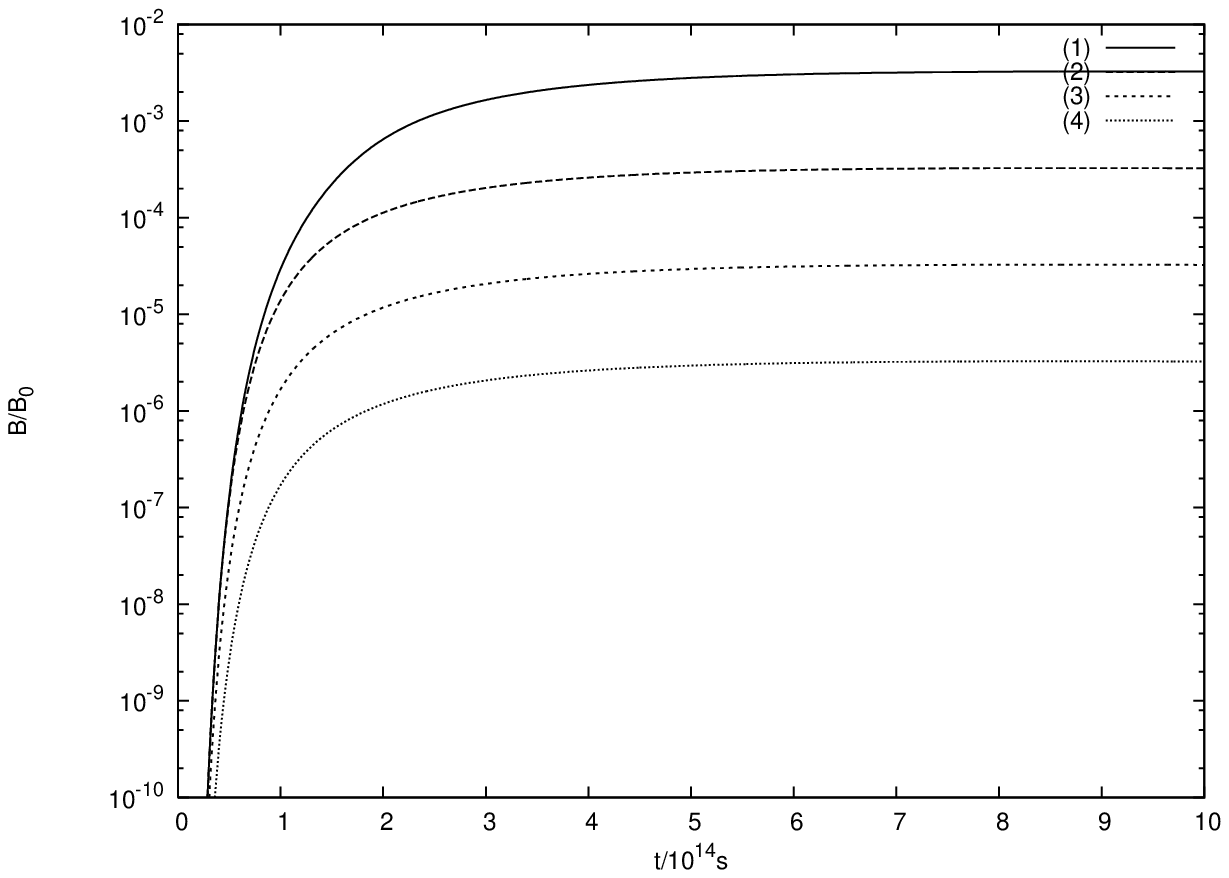}
\caption{The radial magnetic field strength at $r=0.9\,r_{\rm{c}}$
  that results from a constant, uniform, vertical magnetic field
  applied to a DC for (1) $10^{11}\,\rm{s}$, (2) $10^{12}\,\rm{s}$,
  (3) $10^{13}\,\rm{s}$, (4) $10^{14}\,\rm{s}$. The magnetic field
  does not exhibit the same peaks as at $r=0.99\,r_{\rm{c}}$ because the
  saturation of the field takes significantly longer than closer to
  the surface. The field here continues to grow even after the rapid
  decay phase of the field closer to the surface as it continues to
  diffuse inwards. We also see that the strength of the field is
  roughly proportional to the lifetime of the external field.}
\label{applied2}
\end{figure}

\begin{figure}
\centering
\includegraphics[width=0.8\textwidth]{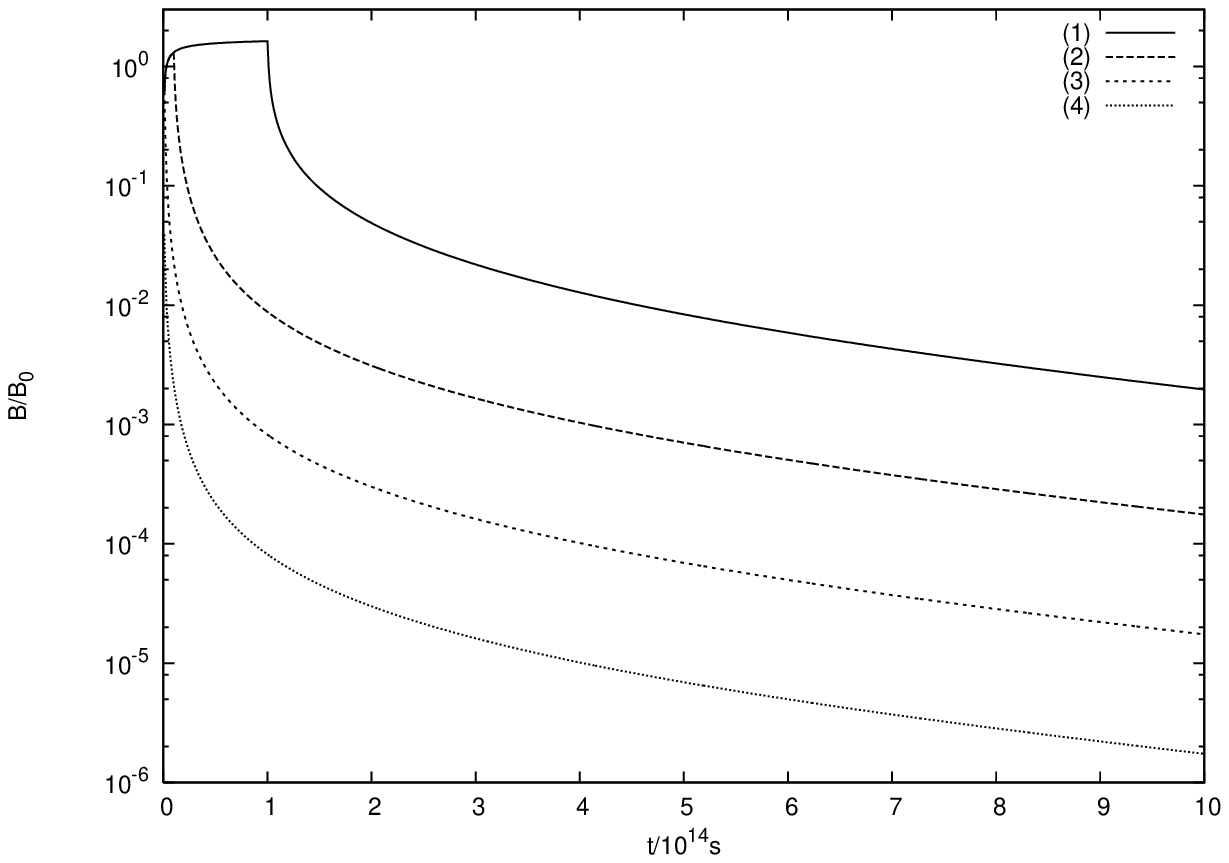}
\caption{The meridional magnetic field strength at
  $r=0.99\,r_{\rm{c}}$ that results from a constant, uniform, vertical
  magnetic field applied to a DC for (1) $10^{11}\,\rm{s}$, (2)
  $10^{12}\,\rm{s}$, (3) $10^{13}\,\rm{s}$, (4)
  $10^{14}\,\rm{s}$. Note the peak of the field strength at the end of
  the phase where the external field is applied.  After the removal of
  the external field there is a period of rapid decay lasting around
  $2\times 10^{14}\,\rm{s}$ after which the field decays much more
  slowly on a timescale of around $10^{15}\,\rm{s}$.  The
  strength of the meridional field is several orders of magnitude
  larger than the radial field.  The strength of the
  field is roughly proportional to the lifetime of the external
  field.}
\label{applied3}
\end{figure}

\begin{figure}
\centering
\includegraphics[width=0.8\textwidth]{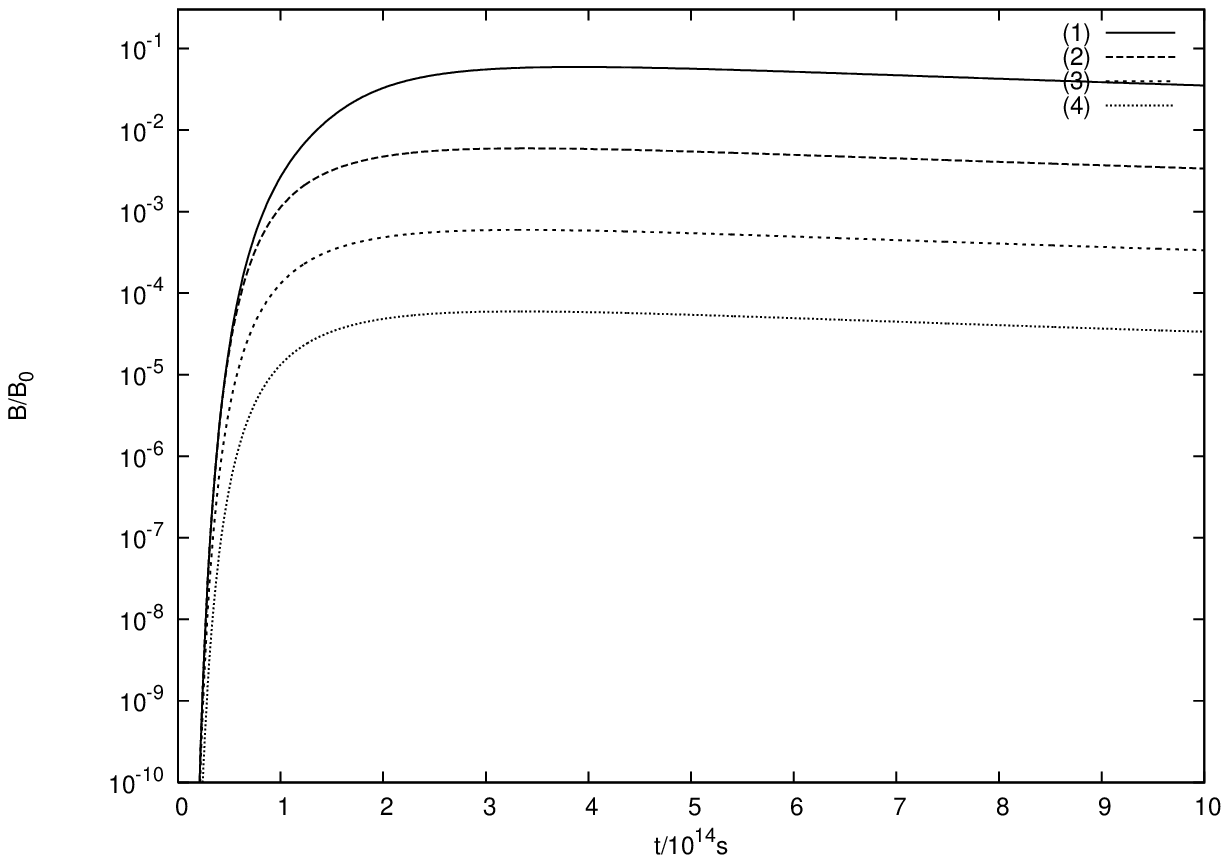}
\caption{The meridional magnetic field strength at $r=0.9\,r_{\rm{c}}$
  that results from a constant, uniform, vertical magnetic field
  applied to a DC for (1) $10^{11}\,\rm{s}$, (2) $10^{12}\,\rm{s}$,
  (3) $10^{13}\,\rm{s}$, (4) $10^{14}\,\rm{s}$. The magnetic field
  does not exhibit the same peaks as at $r=0.99\,r_{\rm{c}}$ because the
  saturation of the field takes significantly longer than closer to
  the surface. The field here continues to grow even after the rapid
  decay phase of the field closer to the surface as it continues to
  diffuse inwards. We also see that the strength of the field is
  roughly proportional to the lifetime of the external field.}
\label{applied4}
\end{figure}

We applied a constant external field of $\textbf{\em{B}}_0$ to a DC of
radius $0.01\,\rm{R_{\odot}}$, mass $0.6\,\rm{M_{\odot}}$ with
diffusivity profile determined by a polytropic index of $3/2$ and
temperature $10^5\,\rm{K}$. The equations for electrical
conductivity from \citet{wendell1987} are not strongly
temperature dependent in the highly degenerate density regime. As
such, the diffusivity profile for WDs of temperatures varying between
$10^3\,$K and $10^7\,$K only show noticeable differences at the
surface. This has no significant effect on the evolution of the DC
field in our model. The magnetic field of the WD scales linearly with
the applied field. We tested the cases where the field was applied for
$10^{11}\,\rm{s},10^{12}\,\rm{s},10^{13}\,\rm{s}$ and
$10^{14}\,\rm{s}$. The radial and meridional fields at
$r=0.99r_{\rm{c}}$ and $r=0.9\,r_{\rm{c}}$ are shown in figures
\ref{applied} to \ref{applied4}. For $r=0.99\,r_{\rm{c}}$ we note the
two distinct phases of the solution. First there is the initial growth
phase in the presence of the external field. Once the external field
is removed the field strength peaks and begins to decay. The strength
of the field that remains after the removal of the external field
increases in proportion to the lifetime of the external
field. Following the removal of the external field, the surface field
of the exposed WD decays rapidly for around $10^{14}\,\rm{s}$ before
continuing to decay exponentially with a characteristic timescale of
around $10^{15}\,\rm{s}$. Whilst this rate of decay is too
rapid to explain the existence of long-lived WD magnetic fields, our
simulations have not included WD cooling. The results of
\citet{wendell1987} indicate that cooling causes the diffusivity to
decrease and significantly extends the decay time scale for the
magnetic field. This effectively freezes the field.  If we suppose
the magnetic field is frozen after $2\times10^{14}\,\rm{s}$ when the
field has relaxed to its post CE state then the residual magnetic
field strength produced by an external field of strength
$B_{\rm{ext}}$ with lifetime $t_{\rm{CE}}$, taking into account both
radial and meridional components, is approximately

\begin{equation}
B_{\rm{res}}=B_{\rm{ext}}\frac{5\,t_{\rm{CE}}\times10^{-16}}{\,\rm{s}}.
\label{res}
\end{equation}

From the results at $r=0.9\,r_{\rm{c}}$ we see that the field
continues to diffuse inwards after the removal of the external field
until it reaches saturation point. This is because the diffusivity
decreases towards the interior of the WD. Once the field reaches a
certain point, the diffusion becomes so slow that it is effectively
halted. This is extremely important if we wish to build strong surface
fields because it prevents any further redistribution of magnetic
energy. If we look at the magnetic field further into the star we see
that the field doesn't penetrate any deeper than $r=0.5\,r_{\rm{c}}$
after $10^{15}\,\rm{s}$ and the field inside $r=0.9\,r_{\rm{c}}$ is
extremely weak compared to the surface field.

\subsection{Effect of randomly varying magnetic field orientation}

If the direction of the applied field is not constant, as we might
expect from a dynamo driven field, then typically the final field
strength is reduced by some factor based on the degree of
variation. For a spatially constant conductivity with oscillating
boundary conditions we may solve the induction equation
analytically. This gives us some insight into the mechanisms
preventing the build up of strong fields in the case of a rapidly
varying external field. We also simulated the change in orientation
numerically by taking the field generated by applying a magnetic field
for a short time in a single orientation and then taking the sum of
the same field rotated at random angles at each timestep up until the
dispersion of the CE.

\subsubsection{Uniform diffusivity DC}
\label{Qparameter}

In the case where $\eta_r$ is a constant we are able to solve for $R$
analytically as in section \ref{equations}. In this case,
$\eta(t)=\eta_r\eta_t(t)$ and equation~(\ref{Helmholtz}) becomes the
Helmholtz equation and we seek solutions in the form of spherical
Bessel functions. We find that $R\propto
j_1\left(\left(\rm{i}\gamma\right)^{\frac{1}{2}}r\right)$. Then from
equations (\ref{boundary2}) and (\ref{S}) we find

\begin{equation}
S=\frac{3}{4}\int_{-\infty}^{\infty}\hat{B}_z(\gamma)\left(\frac{j_1\left(\left(\rm{i}\gamma\right)^{\frac{1}{2}}r\right)}{j_1\left(\left(\rm{i}\gamma\right)^\frac{1}{2}r_c\right)+\frac{1}{2}\left(\rm{i}\gamma\right)^\frac{1}{2}j_1'\left(\left(\rm{i}\gamma\right)^\frac{1}{2}r_c\right)}\right)e^{\rm{i} \gamma H}\,d\gamma \,\cos\theta.
\end{equation}

\noindent Consider the strength of the field generated at the surface of the DC by each $\gamma$-mode. We define $Q(\gamma)$ by

\begin{equation}
Q(\gamma)=Re\left[\frac{j_1\left(\left(\rm{i}\gamma\right)^{\frac{1}{2}}r_c\right)}{j_1\left(\left(\rm{i}\gamma\right)^\frac{1}{2}r_c\right)+\frac{1}{2}\left(\rm{i}\gamma\right)^\frac{1}{2}j_1'\left(\left(\rm{i}\gamma\right)^\frac{1}{2}r_c\right)}\right].
\end{equation}

\begin{figure}
\centering
\includegraphics[width=0.6\textwidth]{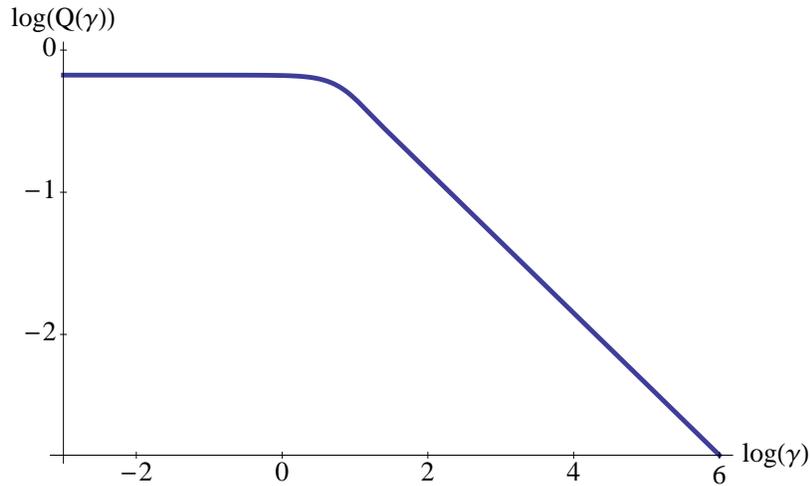}
\caption{Behaviour of the transfer efficiency factor, $Q(\gamma)$ as described in section \ref{Qparameter}.}
\label{Q}
\end{figure}

\noindent We shall refer to $Q(\gamma)$ as the transfer efficiency
because it represents roughly the radial magnetic flux across the
surface of the DC relative to the imposed field. Fig.~\ref{Q}
shows how $Q(\gamma)$ varies. We see that for $\gamma>1$, $Q(\gamma)$
falls off approximately as $\gamma^{-\frac{1}{2}}$. So the higher
$\gamma$ modes of the system are more effectively suppressed but the
fall off is slow. This means that the radial magnetic field at the
surface of the star for very large $\gamma$ approaches $0$. In this
case the boundary conditions are matched by the dipolar field which
cancels out the imposed field at the surface of the DC. We can see
this by evaluating the term in the integrand of equation~\ref{dipole}
at $r=r_{\rm{c}}$ for a specific value of $\gamma$. The term tends to
$-2$ as $\gamma\to\infty$.  This is exactly what is required to cancel
the imposed field.

If we consider a system with constant diffusivity $\eta$, then $H=\eta
t$. If we then take $B_z(t)\propto \cos(\alpha t)$ it is easy to show
that the transfer efficiency is $Q(\frac{\alpha}{\eta})$. We interpret
this by recognising that, if the rate of oscillation is too high
compared to the diffusivity each time the external field switches,
most of the field generated by the previous oscillation is cancelled
out. If the diffusivity is high enough, the oscillations are less
effectively cancelled and a residual field is able to build up.

\begin{figure}
\centering
\includegraphics[width=0.8\textwidth]{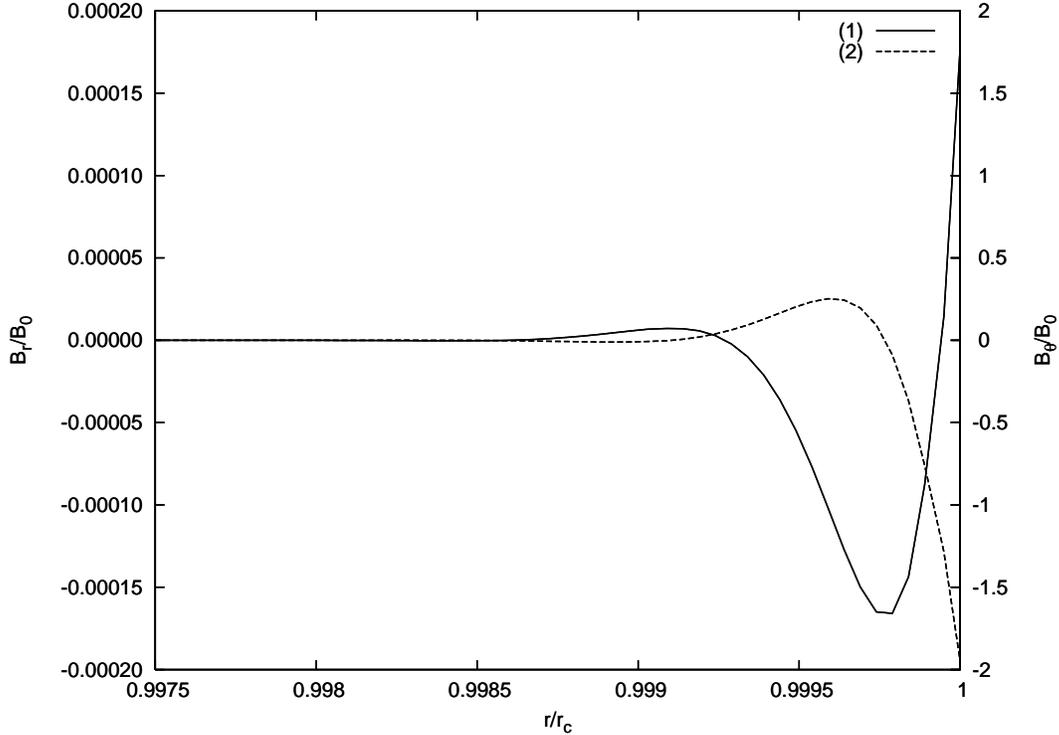}
\caption{Form of the induced field within the DC with a spatially
  uniform magnetic diffusivity $\eta=221\rm{cm}^2\,\rm{s}^{-1}$ with
  variation frequency $\alpha=2\times10^{-8}\,\rm{s^{-1}}$ for an
  imposed field of $B_0$. (1) Radial field ($B_{\rm{r}}$), (2)
  Meridional field ($B_{\theta}$). We see from the horizontal axis
  that the field is confined to the most outer regions of the DC. This
  is because the rapid variation of the external field causes any
  field that is generated at the surface to smooth out to zero as it
  diffuses inwards. We also see that the meridional field is several
  orders of magnitude stronger than the radial field.}
\label{fieldfig}
\end{figure}

Now let us consider the radial form of the induced field. From our
discussion above we expect that the field is functionally similar to
${\rm Re}[j_1({\rm i}{\alpha}/{\eta})^{1/2}r]$. Fig.~\ref{fieldfig} shows
the form of the field that results from this solution (equations
\ref{uniform}, \ref{boundary} and \ref{field}). We have used the
parameters given above and an imposed field strength of $B_0$.  So,
although the transfer efficiency is very low and the radial field is
suppressed, the strong $R$ gradient produces an internal field
parallel to the surface which is comparable in magnitude to the
imposed field but decays extremely rapidly with depth. The thickness
of this layer behaves asymptotically as
$(\frac{\alpha}{\eta})^{-1/2}$. We note that this is the same
behaviour as the efficiency factor $Q(\gamma)$. This is reasonable
because the efficiency of transfer of magnetic energy from the
external field to the DC should scale roughly in proportion to how far
the field can penetrate, at least for shallow layers.

Because we are proposing a model in which the magnetic field of a common envelope surrounding a binary system induces the magnetic field of the DC we choose a value of $\alpha$ to reflect this. First consider a far-field varying on a time scale similar to the orbital period of the binary system. We determine this time scale for a $0.6\,M_{\odot}$ DC with a typical red giant companion of mass $6\,M_{\odot}$ and radius $400\,R_{\odot}$ which has filled its Roche Lobe. This gives us an orbital time scale of $t_{\rm{orbit}}\approx 3.5\times 10^8\,\rm{s}$ and so we choose $\alpha=\frac{2\pi}{t_{\rm{orbit}}}\approx 2\times 10^{-8}\,\rm{s}^{-1}$. This gives a transfer efficiency of $Q(\frac{\alpha}{\eta})\approx 2\times 10^{-4}$. Alternatively as we suggested earlier, the magnetic dynamo is likely to be stronger when the orbit of the binary has decayed. If we apply the same analysis with $t_{\rm{orbit}}=1\,\rm{day}$ then $Q\approx4\times10^{-6}$. In either case, the field is still confined to a very thin boundary layer and is almost entirely meridional.

\subsubsection{Post CE evolution of the magnetic field}

Following the initial generation of the magnetic field within the DC
we foresee two possible cases for the subsequent evolution. Either the
CE is dispersed on a time scale shorter than the variation of magnetic
diffusivity or the diffusivity varies by a significant factor over the
lifetime of the CE. In the absence of a magnetic field, the lifetime
of the CE phase is estimated on the order of $10^3$yr
\citep{taam1978}.  The time scale for orbital decay as a result of a
dynamo driven wind is estimated to be a few Myr \citep{regos1995}.  If
energy from orbital decay is transformed to magnetic energy rather
than unbinding the envelope then the lifetime of the CE phase is
prolonged though from the energy considerations of section
\ref{energy} the lifetime is likely to lie much closer to the lower
bound. Depending on the degree to which this happens the lifetime of
the dynamo may lie anywhere between these two bounds. Typically the
diffusivity varies on a time scale of $t_\eta\approx
10^{16}\,\rm{s}\approx3\times 10^{10}\,\rm{yr}$, much longer than the
probable lifetime of the CE.  However our understanding of CE
evolution is sufficiently poor that these estimates may bear little
resemblance to reality.

The decay of an unsupported magnetic field occurs on a time scale
$t_{B}\approx\beta\frac{r_{\rm{c}}^2}{\eta}$ where $\beta$ is a
constant related to the field geometry For an unsupported field,
$\beta=\pi^2$. In the case where $\eta=221\,\rm{cm}^2\,\rm{s}^{-1}$
which is a good approximation for the surface layers of a DC and
$r_{\rm{c}}=0.01\,R_{\odot}$ we get $t_{B}\approx
2\beta\times10^{15}\,\rm{s}$.  \citet{wendell1987}
showed that the magnetic field of a WD without an imposed field has a
decay time which is much longer than the evolutionary age of the WD
over its whole lifetime and so the magnetic field is essentially
frozen into the WD. We observe that, in our setup where the
internal field is supported by a rapidly oscillating external field,
the strong confinement of the field to the outermost layers of the
DC produces strong gradients which lead to rapid magnetic
diffusion. Once the field is removed, the system rapidly relaxes to
the solution given by zero external field. The field then decays on an
ohmic time scale of $t_{B}$ with $\beta=\pi^2$ \citep{proctor1994}.

\begin{figure}
\centering
\includegraphics[width=0.8\textwidth]{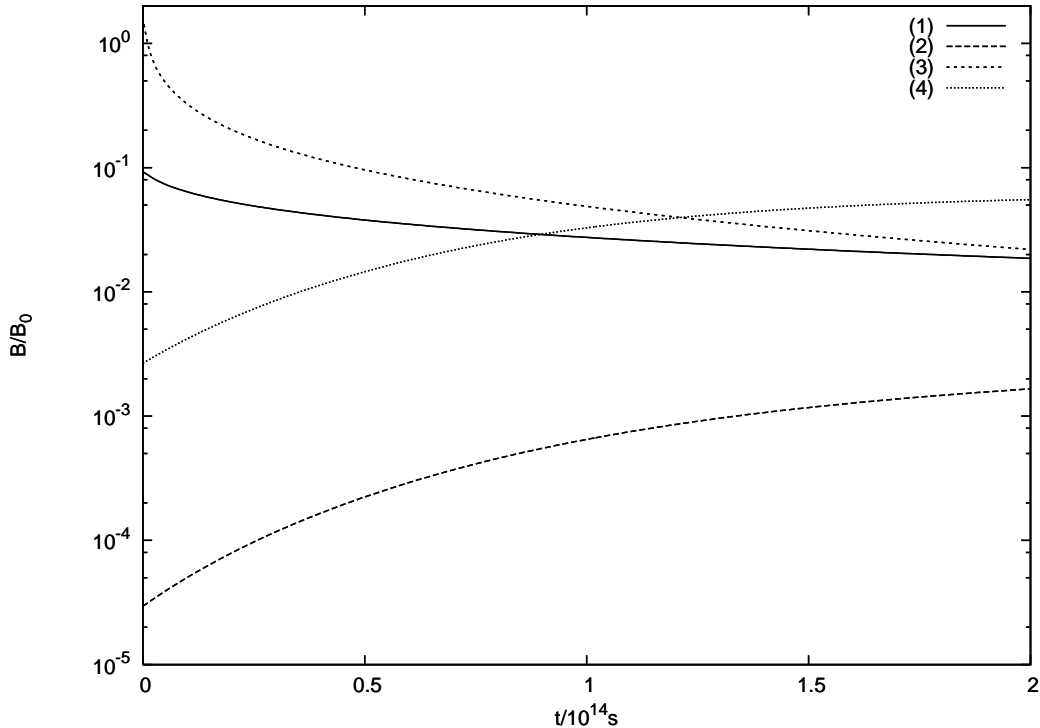}
\caption{Decay of the magnetic field of a WD upon removal of a
  constant external magnetic field applied for $10^{14}\,\rm{s}$.(1)
  $B_{\rm{r}}(r=0.99\,r_{\rm{c}})$ (2)
  $B_{\rm{r}}(r=0.9\,r_{\rm{c}})$, (3)
  $B_{\theta}(r=0.99\,r_{\rm{c}})$, (4)
  $B_{\theta}(r=0.9\,r_{\rm{c}})$.  Although the surface
  meridional field is initially much stronger than the radial field it
  decays much faster until the two are roughly equal
  $2\times10^{14}\,\rm{s}$ after the external field is removed. We
  also see that, even after the rapid initial decay of the surface
  field seems to have finished, the field at $0.9\,r_0$ is still
  growing as the field diffuses inwards.}
\label{decay}
\end{figure}

Fig.~\ref{decay} shows the evolution of the field produced by a DC
with diffusivity profile determined from a polytropic structure. The
field strengths of the radial and meridional fields are shown at
$r/r_{\rm{c}}=0.99$ and $0.9$. We see that, once the external field
vanishes, the field spreads slowly inwards. The field was calculated
below $r/r_{\rm{c}}=0.9$ but even by the end of the simulation, no
field had reached as deep as $r/r_{\rm{c}}=0.5$. The surface magnetic
field decays by about a factor of $10^2$ as it relaxes owing to the
redistribution of the magnetic energy over the radius. This occurs on
a time scale of around $2\times10^{13}\,\rm{s}$. After the initial
rapid decay, the surface field continues to decay exponentially on a
much longer timescale of around $10^{14}\,\rm{s}$. This is faster than
predicted analytically but as Fig.~\ref{decay} shows the field at
$r/r_{\rm{c}}=0.9$ is still growing so the system has not fully
relaxed. It is likely that the decay timescale grows as the field
moves towards spatial equilibrium. WD cooling which has not been
included here would also prevent additional decay.

We have neglected secondary effects that are produced as a result of the finite time the magnetic dynamo of the CE takes to decay. We expect though, given that the lifetime of the CE is estimated to be significantly shorter than the relaxation time for the field, the magnetic dynamo is likely to decay on an even shorter time scale. Thus the effect on the solution should be negligible.

\subsubsection{Numerically evolved field with random orientation}

\begin{table}
\centering
\begin{tabular*}{0.5\textwidth}{@{\extracolsep{\fill}}c c c c}
\hline
$t{\rm{dyn}}/\,\rm{s}$&$B_{\rm{R}}/\rm{G}$&$B_{\rm{M}}/\rm{G}$&$B_{\rm{T}}/\rm{G}$\\
\hline
\multicolumn{4}{c}{No preferred orientation}\\
\hline
$10^6$&$1.9\times10^{-6}$&$-1.8\times10^{-6}$&$1.3\times10^{-5}$\\
$10^7$&$-3.3\times10^{-6}$&$4.0\times10^{-7}$&$-4.1\times10^{-6}$\\
$10^8$&$-3.6\times10^{-6}$&$-2.6\times10^{-6}$&$-1.6\times10^{-6}$\\
$10^9$&$4.2\times10^{-5}$&$-1.6\times10^{-5}$&$3.9\times10^{-6}$\\
\hline
\multicolumn{4}{c}{Half-plane restricted orientation}\\
\hline
$10^6$&$2.3\times10^{-3}$&$-4.2\times10^{-6}$&$-1.1\times10^{-5}$\\
$10^7$&$2.9\times10^{-3}$&$-2.5\times10^{-6}$&$3.8\times10^{-6}$\\
$10^8$&$1.9\times10^{-3}$&$-9.2\times10^{-6}$&$8.7\times10^{-6}$\\
$10^9$&$2.0\times10^{-3}$&$1.6\times10^{-5}$&$1.4\times10^{-5}$\\
\hline
\end{tabular*}
\caption{Residual surface magnetic field strengths generated by an external field of unit strength which reorients at intervals of $t_{\rm{dyn}}$. $B_{\rm{R}}$ is the radial field, $B_{\rm{M}}$ is the meridional field and $B_{\rm{T}}$ is the toroidal field. The external field lasts for a total of $10^{13}\,\rm{s}$ and the residual field is taken at $2\times10^{14}\,\rm{s}$}
\label{orient}
\end{table}

In order to simulate a randomly varying external field in a more
three-dimensional sense we considered a supposition of multiple copies
of a base field rotated into random orientations applied in
sequence. The base field was generated by applying a constant vertical
field for a range of times much shorter than the lifetime of the
CE. Such a superposition is valid because the induction equation is
linear.  So if the field generated by applying a constant vertical
magnetic field for time $\delta t$ is $\tilde{{\textbf{\em{B}}}}(t)$ then
the total field is

\begin{equation}
{\textbf{\em{B}}}(t)=\sum_{k=1}^N{\textbf{\em{n}}}_k\times\tilde{{\textbf{\em{B}}}}(t+k\delta t),
\end{equation}

\noindent where ${\textbf{\em{n}}}_i$ are randomly generated vectors and
$N\delta t$ is the duration of the magnetic dynamo. The results are
shown in Table~\ref{orient}.

We find that in the case of a purely random orientation, the residual
field is smaller than the constant external field case by a factor of
around $1000$. This is to be expected because, on average, each field
orientation is opposed by one pointing in the opposite direction. The
fluctuations arise from the random nature of the field orientation and
the different times at which each field is applied. It seems certain
that the magnetic field strength that would be needed to generate
observed magnetic fields would be extremely large in this case.

\subsubsection{Effect of introducing a preferred direction to the external field}

Due to the nature of the proposed magnetic dynamo it seems reasonable
to suggest that there may be a preferred direction to the magnetic
field. If the field is generated by perturbations in the CE owing to
the orbital motion of the DC and its companion, the magnetic field
lines may show a tendency to align with the orbit. This
  contrasts with the dynamo mechanism in accretion discs which is a
  result of the magneto-rotational instability producing eddies that
  have no preferred direction. Therefore we restrict the magnetic
field vector to lie within a hemisphere. The results are also included
in Table~\ref{orient}. We find that the residual radial field can now
reach strengths close to those produced by the constant external
field. This does not apply to the meridional field or the toroidal
field because different orientations may still oppose each other and
give typically weak field strengths.

If we now vary the CE lifetimes (Table~\ref{orient2}) we see that in
the case of purely random orientations, the overall field strength is
largely unaffected by changes in the CE lifetime. In the case where
the field has preferred alignment, the residual radial field strength
increases roughly linearly with CE lifetime as in the case of constant
external field. It is also able to produce fields with similar field
strength to the uni-directional case. This suggests that while dynamo
action may result in fluctuating magnetic field orientations with
time, the residual field strength can be maintained provided that the
dynamo has a preferred direction on average.

\begin{table}
\centering
\begin{tabular*}{0.5\textwidth}{@{\extracolsep{\fill}}c c c c}
\hline
$t_{\rm{CE}}/\rm{s}$&$B_{\rm{R}}/B_0$&$B_{\rm{M}}/B_0$&$B_{\rm{T}}/B_0$\\
\hline
\multicolumn{4}{c}{No preferred orientation}\\
\hline
$10^{11}$&$4.7\times10^{-6}$&$1.8\times10^{-6}$&$4.0\times10^{-7}$\\
$10^{12}$&$3.2\times10^{-6}$&$-4.8\times10^{-6}$&$-8.5\times10^{-7}$\\
$10^{13}$&$4.2\times10^{-5}$&$-1.6\times10^{-5}$&$3.9\times10^{-6}$\\
$10^{14}$&$-6.7\times10^{-5}$&$-3.3\times10^{-5}$&$-4.9\times10^{-5}$\\
\hline
\multicolumn{4}{c}{Half-plane restricted orientation}\\
\hline
$10^{11}$&$1.9\times10^{-5}$&$3.0\times10^{-7}$&$2.8\times10^{-6}$\\
$10^{12}$&$2.0\times10^{-4}$&$1.1\times10^{-6}$&$1.5\times10^{-5}$\\
$10^{13}$&$2.0\times10^{-3}$&$-2.3\times10^{-5}$&$1.3\times10^{-5}$\\
$10^{14}$&$3.1\times10^{-2}$&$-1.2\times10^{-4}$&$9.5\times10^{-5}$\\
\hline
\end{tabular*}
\caption{Residual surface magnetic field strengths generated by an
  external field of strength $B_0$ which reorients at intervals of
  $10^9\,\rm{s}$. $B_{\rm{R}}$ is the radial field, $B_{\rm{M}}$ is
  the meridional field and $B_{\rm{T}}$ is the toroidal field. The
  external field lasts for $t_{\rm{CE}}$ and the
  residual field is taken at $2\times10^{14}\,\rm{s}$}
\label{orient2}
\end{table}

\section{Conclusions \& Discussion}

The statistical occurrence of magnetic WDs in close interacting binary
systems suggests that the origin of their strong fields is a product
of some feature of their binary evolution. Indeed, the argument that
the fields originate through flux conservation during the collapse of
Ap/Bp stars does not explain the higher frequency in interacting
binaries. By building on the proposal that the magnetic field is
generated by dynamo action within a common convective envelope we have
shown that strong fields may be transferred to the DC.  These fields
can then be preserved upon dissipation of the envelope.

Following the dissipation of the CE, we have found that the system
rapidly relaxes on a timescale of about $7\,\rm{Myr}$ to the state we
would anticipate given no external field. Although there is no
significant dissipation of magnetic energy during this time, the
redistribution of magnetic flux towards the interior of the WD results
in significant decay of the surface field by around a factor of at
least $10-100$. Further decay is prevented by the increasing
electrical conductivity towards the interior of the WD. This prevents
field diffusing beyond around $r=0.9r_{\rm{c}}$. Once the field has
relaxed to its new configuration it continues to decay but on a
timescale much longer than the time scale for cooling of the WD
so we expect any further loss of field strength to be minimal.

The maximal rate of transfer of field energy from the CE to the DC
occurs when the field is kept fixed. In this case the residual
strength of the field following dissipation of the CE is linearly
related to the lifetime of the CE. If we combine equations (\ref{me})
and (\ref{res}) we find

\begin{equation}
M=5.7\times10^{35}\left(\frac{B_{\rm{res}}}{10^7\,\rm{G}}\right)^2\left(\frac{a}{0.01\,\rm{au}}\right)\left(\frac{r_{\rm{I}}}{0.01R_{\odot}}\right)^2\left(\frac{t_{\rm{CE}}}{2\times10^{15}\,\rm{s}}\right)^{-2}\,\rm{J}.
\end{equation}

\noindent If we assume that all of the energy released through orbital
decay is transformed into magnetic energy then this requires an
envelope lifetime of
$1.1\times10^{12}\,$s or $3.6\times10^4\,$yr to produce a
$10^7\,$G magnetic field. This lifetime is extended if there is
variation in the direction of the field produced by the dynamo or
there are energy sinks other than the magnetic field. Because the
required envelope lifetime scales linearly with $B_{\rm{res}}$, weaker
fields could be produced on a much shorter time
scale. Energetically there is nothing to prevent formation of
a $10^9\,$G MWD. As shown, the final field strength scales
proportionally with the CE lifetime. If the dynamo is confined to a
smaller region or the CE lifetime is extended then this strength of
field could be produced. Scenarios this extreme seem unlikely and this is
complemented by the rarity of MWDs with such strong magnetic
fields. We do not rule out the possibility that the strongest fields
may be produced by the merger of a white dwarf and the core of a giant
star which have both been strongly magnetized during CE
phases.

We have also shown that the production of sufficiently strong fields
is almost certainly dependent on some preferred orientation of the
magnetic dynamo. In the case where there is no preferred direction the
DC magnetic field is confined to a layer of thickness about
$(\frac{\alpha}{\eta})^{-1/2}$, where $\alpha$ is the frequency for
field variation, and is almost entirely meridional. Despite the
generation of strong surface fields during the CE phase in this case,
the total magnetic energy of the DC field is constrained by the depth
of the magnetic layer. Consequently once the CE disperses and the
field of the exposed WD begins to diffuse inwards, the majority of the
field strength is lost. This result is confirmed by the numerical
simulations. Typically a field strength of $10^{-5}$ relative to the
average magnetic field strength of the dynamo is the largest possible
residual field. This may be sufficient to produce WDs with weaker
fields but the formation of the strongest WD magnetic fields would
require far more energy than the system could provide. In the case
where the field has some preferred orientation, the proportion of
field retained is increased up to a few per cent depending on the
lifetime of the CE dynamo.

Whilst the arguments we have presented here suggest that the origin of
the fields of MWDs may be the result of dynamo action in CEs of
closely interacting binaries, there are still many uncertainties. Most
of these are the result of insufficient understanding of the formation
and evolution of CEs. The field structures we have used to construct
our models are simple but complex field geometries are
observed in WDs such as Feige~7 and KPD 0253+5052. If CE dynamos are
responsible for the origins of MWDs, they should support complex
geometries. Further progress depends on a better understanding of the
physical processes and energy transfer within the CE.

\section*{Acknowledgements}

This work was supported by the STFC.

\label{lastpage}
\end{document}